\documentclass[aps, prl, twocolumn]{revtex4}
\usepackage{amsmath}
\usepackage{amssymb}
\usepackage{amsthm}
\usepackage{xspace}
\usepackage{graphicx}
\usepackage{microtype}
\usepackage{paralist}
\usepackage{nicefrac}
\usepackage{adjustbox}
\usepackage{hyperref}
\usepackage[usenames,dvipsnames,table]{xcolor}

\newcommand{\inner}[2]{\langle #1, #2 \rangle}

\newcommand{\calA}{\mathcal{A}}
\newcommand{\calB}{\mathcal{B}}
\newcommand{\abort}{\mathrm{abort}}
\newcommand{\prob}[2]{\mathsf{Prob}(#1,#2)}

\begin{document}

\title{\vspace{-1cm}On the impossibility of coin-flipping in generalized probabilistic theories\\via discretizations of semi-infinite programs}

\author{Jamie Sikora}
\affiliation{Perimeter Institute for Theoretical Physics, Waterloo, Ontario, Canada, N2L 2Y5}

\author{John H. Selby}
\affiliation{Perimeter Institute for Theoretical Physics, Waterloo, Ontario, Canada, N2L 2Y5}

\date{\today}

\begin{abstract}
Coin-flipping is a fundamental cryptographic task where a spatially separated Alice and Bob wish to generate a fair coin-flip over a communication channel.
It is known that ideal coin-flipping is impossible in both classical and quantum theory.
In this work, we give a short proof that it is also impossible in generalized probabilistic theories under the Generalized No-Restriction Hypothesis.
Our proof relies crucially on a formulation of cheating strategies as semi-\emph{infinite} programs, i.e., cone programs with infinitely many constraints.
This introduces a new formalism which may be of independent interest to the quantum community.
\end{abstract}


\maketitle

In this paper we consider the possibility of cryptography in theories more general than quantum or classical theory.
One may ask why this is a worthwhile endeavour, and for this we give several reasons.
The first reason is to future-proof current results which is important in the context of cryptography.
While developing quantum cryptography and computation, the community quickly came to realize that classical cryptography results need to be reevaluated for the new quantum era.
Since results in quantum cryptography typically rely on the validity of quantum mechanics being a faithful description of nature, these too all have to reevaluated if quantum theory is one day superseded by a new theory, regardless of how minor or radical the departure from quantum mechanics is.
Another reason is to gain a better understanding
of results in quantum theory.
For instance, it is insightful to sit back and think about what parts of quantum theory were needed to prove a result.
Did we require entanglement?
Were we just assuming these states are in superposition?
Can we reprove this only assuming the No-Signalling Principle?
By answering such questions, we gain a better understanding of quantum mechanics itself as well as the resources necessary for performing particular tasks.

In this and many other works in cryptography, optimization theory is a key ingredient in the analysis.
On a high level, we want to maximize how much someone can ``cheat'' a protocol, whereby it is understood that the inability to cheat translates into security, and vice versa.
The goal is often to design protocols which minimize cheating.
We, however, take the opposite approach in this work and prove a limitation on designing \emph{any} protocol for a particular task, namely coin-flipping, discussed below.

\

\noindent\textit{Coin-flipping---}
Coin-flipping is the cryptographic task where Alice and Bob generate a random bit $b$ over a communication channel such that, when Alice and Bob are honest, both output the same bit $b$ and this bit is uniformly random \cite{Blu81}. Coin-flipping is a primitive that is used mainly for building larger, more sophisticated cryptographic protocols in the two-party setting, and hence an understanding of its properties, along with its security limitations, is important.

More formally the coin-flipping task is as follows. Suppose Alice has a set of strategies (basically, a description of how she interacts with Bob) given by the set $\calA$ and Bob has a set of strategies given by the set $\calB$.
We do not just consider deterministic strategies but also those that occur as the result of some measurement procedure.
We denote the probability of a pair of strategies occurring as $\prob{A}{B}$ which is between $0$ and $1$ for all $A \in \calA$ and $B \in \calB$.

A coin-flipping protocol consists of the following:
\begin{itemize}
\item A triple of strategies for Alice $(A_0, A_1, A_{\abort})$ which correspond to the measurement outcomes of some deterministic strategy $A_{\text{det}}$,
\item A triple of strategies for Bob $(B_0, B_1, B_{\abort})$ which correspond to the measurement outcomes of some deterministic strategy $B_{\text{det}}$,
\end{itemize}
satisfying
\begin{equation} \label{conditions2}
\prob{A_b}{B_b} = 1/2 \; \text{ for } \; b \in \{ 0, 1 \}.
\end{equation}
The conditions above ensure that the protocol behaves as expected, that the bit $b$ is uniform and shared between Alice and Bob.
Ideally, we wish that neither Alice nor Bob can cheat by digressing from protocol and disturbing the conditions given by \eqref{conditions2}.
However, this may not be the case, and as such, we need to measure this disturbance.
The security measure in coin-flipping is given by the amount a dishonest Alice or a dishonest Bob can bias the output distribution away from uniform.
To make this formal, we define the symbols:
\begin{itemize}
\item $P_{\text{Alice},b}^*$ : The maximum probability that dishonest Alice can force honest Bob to accept the outcome $b$.
\item $P_{\text{Bob},b}^*$ : The maximum probability that dishonest Bob can force honest Alice to accept the outcome $b$.
\item $\epsilon$: The bias of the coin-flipping protocol defined as
\begin{equation}
\epsilon := \max \{ P_{\text{Alice},0}^*, P_{\text{Alice},1}^*, P_{\text{Bob},0}^*, P_{\text{Bob},1}^* \} - 1/2.
\end{equation}
\end{itemize}
We wish to design protocols such as to minimize $\epsilon$, with a perfect protocol having $\epsilon = 0$.
In classical and quantum theory, this is known to be impossible~\cite{LC97a, Kit03}.
In this work, we show that under some assumptions on $\calA$ and $\calB$,
$\epsilon$ can be lower bounded by a positive constant, thus showing near-perfect coin-flipping is impossible in any theory satisfying those assumptions.

To study the range of possible $\epsilon$, we need to study the four quantities $P_{\text{Alice},0}^*$, $P_{\text{Alice},1}^*$, $P_{\text{Bob},0}^*$, and $P_{\text{Bob},1}^*$.
Let us first consider $P_{\text{Bob},0}^*$.
We can write this succinctly by the rudimentary optimization problem:
\begin{equation} \label{rudebobNew}
P_{\text{Bob},0}^* = \sup_{B \in \calB} \left\{ \prob{A_0}{B} \right\}.
\end{equation}
This optimization problem exactly captures how much Bob can force Alice to output $0$ maximized over all physical strategies he can perform.
Before studying this problem using optimization theory,  we require a mathematical structure on the quantities involved.
We now discuss such a structure which is given by the study of Generalized Probabilistic Theories.

\

\noindent\textit{Generalized Probabilistic Theories \textup{(GPTs)}---}
To study \eqref{rudebobNew} more generally than quantum and classical theory we require a more general setting for physical theories.
Here we work in the framework of generalized probabilistic theories which formalizes any physical theory with an operational description.
There have been many approaches to GPTs, see, for example, \cite{hardy2001quantum,barrett2007information,Ludwig,davies1970operational,randall1970approach,Piron64,Mackey,chiribella2010probabilistic,hardy2011reformulating} for introductions to these frameworks.
GPTs have been successfully used for studying cryptography \cite{sikora2018simple,selby2018make,lami2018ultimate,barnum2011information,barnum2008nonclassicality,barrett2007information,barrett2005no} and computation \cite{krumm2018quantum,barnum2018oracles,garner2018interferometric,barrett2017computational,lee2016deriving,lee2016bounds,lee2016generalised,lee2015computation,lee2017higher} in theories more general than quantum theory. We, however, do not actually need to introduce the full framework of GPTs for the purposes of this work. Instead, we just consider the structure that any such theory would impose on the sets of strategies for Alice and Bob.

As mentioned above, we do not just want to consider the strategies which occur deterministically, but those which may correspond to obtaining a particular outcome in some experiment.
That is, given a strategy $A \in \calA$ for Alice and a strategy $B \in \calB$ for Bob we obtain a probability $\mathsf{Prob}(A,B)$ that these two strategies jointly occur.
In particular there is always a `zero-strategy' $0 \in \calA$ such that $\mathsf{Prob}(0,B)= 0$ for all $B \in \calB$.
Conceptually, one can think of this as Alice aborting the protocol, or simply not taking part in the first place.

First, we assume that these spaces of strategies are \emph{convex} where we interpret convex combinations as \emph{probabilistic mixtures}. That is, we assume that
\begin{equation}
p A_1 + (1-p) A_2
\end{equation}
is in the set $\calA$ and represents the strategy where  with probability $p$ Alice uses strategy $A_1$ and with probability $1-p$ Alice uses strategy $A_2$.
Given this understanding of the convex structure, the calculated probabilities must satisfy
\begin{equation}
\mathsf{Prob}\left(\sum_i p_i \, A_i, B \right) = \sum_i p_i \, \mathsf{Prob}(A_i, B)
\end{equation}
and similarly for convex combinations of Bob's strategies.
This means that a strategy for Alice induces a linear functional on the space of strategies for Bob (and vice versa).

Rather than working directly with the spaces of strategies $\calA$ and $\calB$ we work with operational equivalence classes of strategies.
We say that two strategies $A_1$ and $A_2$ are operationally equivalent if
\begin{equation}
\mathsf{Prob}(A_1, B) = \mathsf{Prob}(A_2,B), \quad \forall B\in\calB
\end{equation}
and similarly for Bob's strategies.
We denote these equivalence classes as $\tilde{\calA}$ and $\tilde{\calB}$.

Note that our earlier assumptions imply that $\tilde{\calA}$ and $\tilde{\calB}$ are both convex sets in some vector space $V$ which are bounded and have non-empty interior.
Moreover, we assume that the vector space $V$ is finite-dimensional.
This assumption is typically made in the study of GPTs for technical convenience.
It can however be motivated by the idea that in a tomographic characterization of the strategies of Alice, one can only, in practice, perform a finite number of different experiments and therefore we must characterize the strategies by a finite number of probabilities.

Following a standard argument on the representations of linear functionals on finite-dimensional vector spaces, one can show that we can always write probabilities as
\begin{equation}
\mathsf{Prob}(A,B) = \inner{\tilde{A}}{\tilde{B}}.
\end{equation}
From now on we take $\tilde{\calA}$ as the set of Alice's strategies (similarly $\tilde{\calB}$ as the set of Bob's strategies) and hence drop the tildes for convenience as the strategy representation should be clear from context.

We can now rewrite the optimization problem \eqref{rudebobNew} in the form
\begin{equation} \label{rudebob}
P_{\text{Bob},0}^* = \sup_{B \in \calB}
\left\{
\inner{A_0}{B}
\right\}.
\end{equation}
Due to the convex structure of the set $\calB$, this is a  convex optimization problem.
However, since we want to prove general bounds on cheating, we require more structure on the sets $\calA$ and $\calB$ for our analysis.

\

\noindent\textit{A physical assumption---}
Clearly some assumption on the sets $\calA$ and $\calB$ is required to prove anything meaningful.
For example, consider any physical theory and restrict both Alice and Bob to a set of strategies that are
{$\epsilon$-close} to their honest strategies.
This allows us to define a (rather boring) GPT in which ideal coin-flipping is possible up to some small error.
To avoid GPTs with these unnecessary restrictions,
we make the assumption that any mathematically feasible strategy for Bob can be physically realized.

To formally define this lack of restriction for Bob,  we start with defining two important quantities studied in convex analysis.
The \emph{polar set} of the set $C$ is given as
\begin{equation}
C^o := \{ W : \inner{W}{Z} \leq 1, \, \forall Z \in C \}
\end{equation}
and its \emph{dual cone} is given as
\begin{equation}
C^* := \{ W : \inner{W}{Z} \geq 0, \, \forall Z \in C \}.
\end{equation}
Notice we have $\calB \subseteq \calA^* \cap \calA^o$ and $\calA \subseteq \calB^* \cap \calB^o$ because
\emph{every} choice of strategies for Alice and Bob  yields a proper probability.

We can now define our physical assumption.
\quad \\ \quad \\
\noindent
\textbf{Definition 1}.
The \emph{Generalized No-Restriction Hypothesis for Bob} states that $\calB = \calA^* \cap \calA^o$.
\quad \\

To support this assumption, one can argue that if Alice knows that her set of strategies is given as $\calA$ then to be able to guarantee security against Bob she should not make any assumptions about what Bob can do.
In other words, we also maximize over all physical theories, which in this case translates to allowing Bob to have the largest set of strategies as possible.

This is closely related to the (standard) No-Restriction Hypothesis \cite{chiribella2010probabilistic} which is a commonly used assumption in the study of GPTs that can be expressed as the idea that all mathematically possible measurements are physically allowed.
Here we generalize this idea to the level of arbitrary strategies.

\begin{figure}[t]
\includegraphics[width=0.48\textwidth,trim={0 21.5cm 0 0cm},clip]{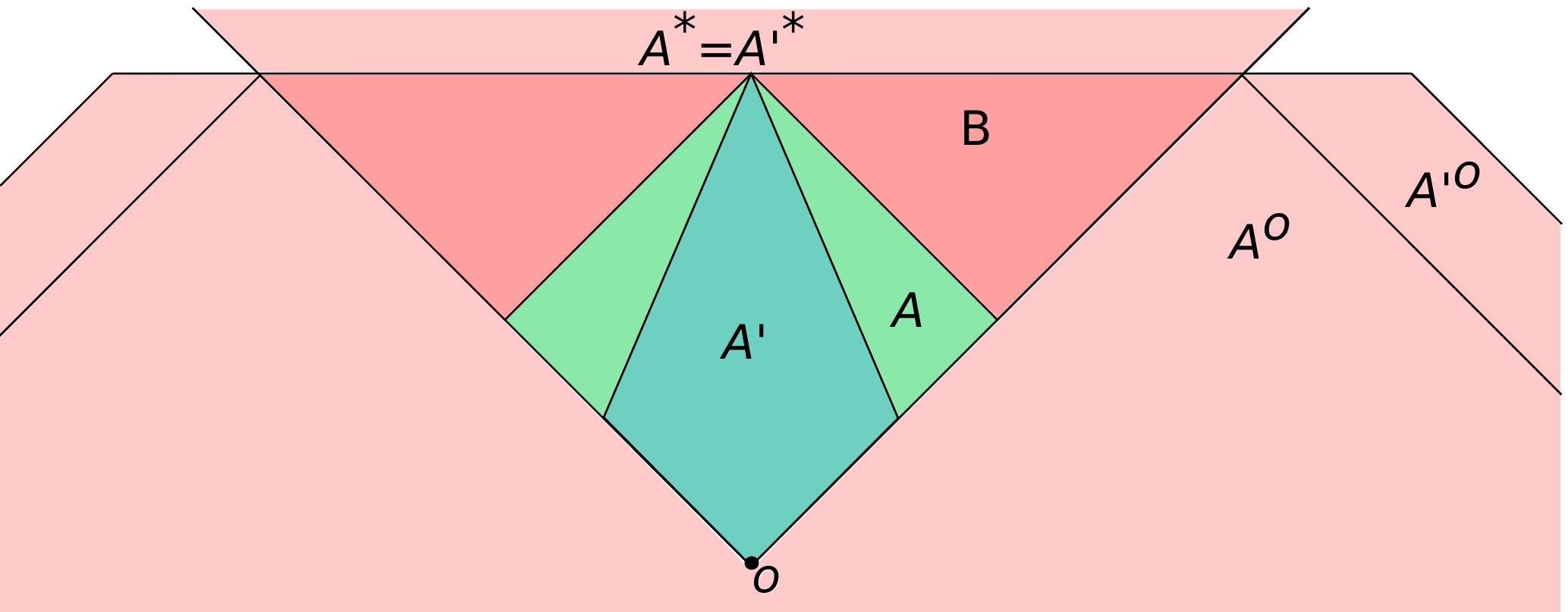}
\caption{
Alice has two strategy sets $A$ and $A'$ corresponding to two different theories.
We see that $B$ is equal to both $A^* \cap A^o$
and $(A')^* \cap (A')^o$ and hence the Generalized No-Restriction Hypothesis for Bob does not imply the same for Alice.
We do have that $A = B^* \cap B^o$, so sometimes the assumption does hold for both Alice and Bob.
}
\label{NoRestrictionPic}
\end{figure}

One could equally well consider Bob's perspective and assume the Generalized No-Restriction Hypothesis for Alice, i.e. $\calA= \calB^*\cap\calB^o$.
Surprisingly these two assumptions are not equivalent,
see Fig.~\ref{NoRestrictionPic} for an example of this fact. However, for the purposes of this work we need to only assume it for one party.
We henceforth assume it for Bob, but by symmetry the following arguments can be adapted to the case where it is assumed instead for Alice.

\vspace{0.25cm}

\noindent\textit{Optimization analysis---}
Under this assumption we can now clean up the optimization problem for Bob \eqref{rudebob} as:
\begin{align}
\!\!\!\! P_{\text{Bob},0}^*
& = \sup_{B \in \calA^* \cap \calA^o} \left\{ \inner{A_0}{B}
\right\} \\
& = \sup_{B \in \calA^*} \left\{ \inner{A_0}{B} : \inner{B}{A} \leq 1, \forall A \in \calA \right\}. \label{nicebob}
\end{align}

This type of optimization problem is called a \emph{semi-infinite program} since the variable $B$ is finite-dimensional but there are infinitely many constraints.  (Note that this class is not the same as the more popular class of optimization problems called semidefinite programs.)
Semi-infinite programming has a rich theory, see for example \cite{Shapiro2009}, although it has yet to be used to study quantum theory or its generalizations, as far as we are aware.

For our needs, it suffices to look at relaxations of $P_{\text{Bob},0}^*$ where we optimize instead using a discretization of the infinite set $\calA$.
To this end, we define a mesh, denoted here as $\calA_{\delta}$, parameterized by a fineness measure $\delta > 0$, such that it has the following properties:
\begin{itemize}
\item $\calA_{\delta}$ is finite, contains a basis for $V$, and is contained in $\calA$;
\item $\forall A \in \calA, \, \exists X \in \calA_{\delta} \, \text{ such that } \,
\| X - A \|_2 \leq \delta$.
\end{itemize}
Note that such a discretization always exists since $\calA$ is bounded.

We now consider the discretized version of this optimization problem defined to optimize using $\calA_{\delta}$ instead, as shown below
\begin{equation} \label{pertbob}
P_{\text{Bob},0}^{\delta} = \sup_{B \in \calA^*} \left\{ \inner{A_0}{B} : \inner{B}{X} \leq 1, \forall X \in \calA_{\delta}
\right\}.
\end{equation}
First note that we have $P_{\text{Bob},0}^* \leq P_{\text{Bob},0}^{\delta}$ since it relaxes \eqref{nicebob} as $\calA_{\delta} \subseteq \calA$.
Furthermore, since there are finitely many constraints, this is a (traditional) cone program making it easier to analyze.
Recently there have been several applications of cone programming to the study of GPTs \cite{selby2018make,sikora2018simple,fiorini2014generalized, JP17,bae2016structure,lami2018ultimate}
and to quantum theory \cite{GSU13, BCJRWY14, LP15, NST16, SW17}.

As expected, as one decreases $\delta$ (the fineness measure of the mesh), we have that $\calA_{\delta}$ becomes a better approximation of the set $\calA$.
In particular, we have the lemma below.

\quad \\
\noindent
\textbf{Lemma 2.}
$\lim_{\delta \to 0^+} P_{\text{Bob},0}^{\delta} = P_{\text{Bob},0}^*$.

\begin{proof}
We first show that the feasible region of $\eqref{pertbob}$ is bounded.
To this end, we define the function
\begin{equation} \label{norm}
f(Y) = \max_{X \in \calA_{\delta}} \{ | \inner{X}{Y} | \}
\end{equation}
which is finite since $\calA_{\delta}$ is finite.
It can be easily checked that this is a norm (since $\calA_{\delta}$ contains a basis) and is bounded for all $B$ satisfying the constraints of \eqref{pertbob}.
Since all norms are equivalent in finite-dimensional vector spaces, we know there exists a $\tau > 0$ such that $\| B \|_2 \leq \tau$ for all $B$ feasible in $\eqref{pertbob}$.

Fix $B$ feasible in $\eqref{pertbob}$ and $A \in \calA$. We now wish to scale $B$ by some constant $c>0$ to ensure $\inner{A}{cB} \leq 1$ (and thus $cB$ is feasible in $\eqref{nicebob}$).
Then for $X \in \calA_{\delta}$ $\delta$-close to $A$,
we have
\begin{align}
\inner{B}{A}
& = \inner{B}{X} + \inner{B}{A-X} \\
& \leq \inner{B}{X} + \| B \|_2 \| A-X \|_2 \\
& \leq 1 + \tau \delta.
\end{align}
Thus, $\dfrac{1}{1 + \tau \delta} B$ is feasible in \eqref{nicebob}. This implies that
\begin{equation} \label{bound}
P_{\text{Bob},0}^* \leq P_{\text{Bob},0}^{\delta} \leq (1 + \tau \delta) \, P_{\text{Bob},0}^*.
\end{equation}
Taking limits finishes the proof.
\end{proof}

We now prove a lower bound on the product of Alice's cheating probability and the relaxation of Bob's cheating probability.
This is the key step in proving our main result which takes advantage of the simplified structure of the relaxed problem.

\quad \\
\noindent
\textbf{Lemma 3.}
${P_{\text{Alice},0}^* \cdot P_{\text{Bob},0}^{\delta} \geq 1/2}$, for all $\delta > 0$.

\begin{proof}
Let $B \in \mathrm{int}(\calB) = \mathrm{int}(\calA^* \cap \calA^o) \subseteq \mathrm{int}(\calA^*)$ which exists since $\calB$ has nonempty interior by construction.
Then $B' := \frac{1}{2} B$ satisfies $B' \in \mathrm{int}(\calA^*)$
and $\inner{B'}{X} < 1$ for all $X \in \calA_{\delta}$.
This is known as a \emph{strictly feasible} solution.
Since $P_{\text{Bob},0}^{\delta}$ is bounded from above by Eq.~\eqref{bound}, the strong duality theorem for cone programming (see, for example, \cite{BV}) states that $P_{\text{Bob},0}^{\delta}$ is equal to
\begin{equation} \label{Bobdual}
\!\!\! \min_{y_X \geq 0} \left\{ \sum_{X \in \calA_{\delta}} y_X : \!
\sum_{X \in \calA_{\delta}} y_X X - A_0 \in (\calA^*)^*
\right\}
\end{equation}
and this problem attains an optimal solution $\{ y'_X \}$. Thus, we have ${P_{\text{Bob},0}^{\delta} = \sum_{X \in \calA_{\delta}} y'_X}$.
Define
\begin{equation}
A :=
\frac{1}{P_{\text{Bob},0}^{\delta}}
\sum_{X \in \calA_{\delta}} y'_X X
=
\sum_{X \in \calA_{\delta}} \left( \frac{y'_X}{\sum_{\tilde{X} \in \calA_{\delta}} y'_{\tilde{X}}} \right) X.
\end{equation}
Notice that $A \in \calA$ by convexity and
$A - \frac{1}{P_{\text{Bob},0}^{\delta}} A_0 \in (\calA^{*})^*$ by the constraints in \eqref{Bobdual}.
Suppose Alice uses $A$ as her strategy to force Bob to accept outcome $0$.
Then we have
\begin{equation}
P_{\text{Alice},0}^* \geq \inner{A}{B_0} \geq \frac{1}{P_{\text{Bob},0}^{\delta}} \inner{A_0}{B_0} = \frac{1}{2 P_{\text{Bob},0}^{\delta}}
\end{equation}
since $B_0 \in \calB \subseteq \calA^*$ and $\inner{A_0}{B_0} = 1/2$ from Eq.~\eqref{conditions2}.
\end{proof}

By combining the two lemmas, we have that ${P_{\text{Alice},0}^* \cdot P_{\text{Bob},0}^{*} \geq 1/2}$, and therefore the maximum of the two probabilities is at least $1/\sqrt{2}$.
This gives the same lower bound on the bias Kitaev gave for the case of quantum theory \cite{Kit03} which was later reproved by Gutoski and Watrous using a representation of quantum strategies~\cite{GW07}.

\quad \\
\noindent
\textbf{Theorem 4.}
Any coin-flipping protocol in a GPT satisfying the \emph{Generalized No-Restriction Hypothesis for Bob} (and/or Alice)
satisfies $\epsilon \geq 1/\sqrt{2} - 1/2 \approx 0.207$.
In particular, either Alice or Bob can force an outcome with probability at least $1/\sqrt{2}$.
\quad \\

Since quantum theory satisfies the Generalized No-Restriction Hypothesis for both Alice and Bob \cite{GW07}, we have another proof that coin-flipping is impossible in quantum theory.

\

\noindent\textit{Discussion---}
What is perhaps unusual about our main result is that we have found a numerical lower bound that holds for any GPT satisfying the Generalized No-Restriction Hypothesis for Alice and/or Bob.
Typically results in the study of GPTs either show something is possible or impossible, or consider a specific GPT (whose structure can be exploited).
This is relevant for cryptographic purposes as well.
If our result was simply saying that perfect coin-flipping is impossible, then this does not rule out the existence of protocols with small bias, which would be enough for all intents and purposes.
Theorem~4 says that near perfect protocols cannot  exist either.
Moreover, the constant lower bound shows that the security of coin-flipping protocols cannot be boosted in the sense that a protocol with bias $\epsilon < 1/2$ cannot be used in a composition to reduce the bias arbitrarily close to $0$.

The main technique in this work is our treatment of semi-infinite programs, in particular, how we discretized them into cone programs.
We hope that our use of semi-infinite programs will raise awareness of this formalism for future uses in quantum theory and physics by breaking roadblocks when formulating difficult problems as optimization problems.

\

\noindent\textit{Future work---}
This bound on coin-flipping is (asymptotically) achievable in quantum theory using a protocol which is classical apart from quantum subroutines~\cite{CK09}.
This quantum subroutine is a black-box implementation of quantum \emph{weak} coin-flipping--a similarly defined task but with less stringent security requirements.
The history of finding the best quantum weak coin-flipping protocol culminated in the work of Mochon \cite{Moc07}.
This unpublished paper is $80$ pages long and, even though it has been simplified \cite{ACGKM15} (see also \cite{NST15}), is still not well understood.
(Recent progress has been made however in the work~\cite{ARW18}.)
Mochon's work relies on point games (developed by Kitaev), a notion which is dual, in a sense, to protocols (specified in this work as the pair of triples $((A_0, A_1, A_{\abort}),(B_0, B_1, B_{\abort}))$.
Even though point games are mysterious in the context of quantum theory, perhaps our generalization to the framework of GPTs will shed light.
In fact, there is one immediate similarity to this work.
A major step in Mochon's proof is the reduction from time-dependent point games to time-independent point games.
This, in a nutshell, strips away all the `time-dependent' information of the protocol.
Our framework and proof, on the other hand, completely strips away all notion of time as it does not explicitly rely on the round-to-round strategy descriptions, and thus might make this point game reduction simpler, or even trivial.

In short, if one were to develop GPT weak coin-flipping protocols with small bias, then the lower bound presented in this work might be achievable by imitating the quantum protocol.
It would be interesting to see which GPTs allow for secure weak coin-flipping, whether it is proved using point games, semi-infinite programming, or another yet-to-be-discovered method.

\medskip
\begin{acknowledgments}
\emph{Acknowledgements--}
We thank Martin Pl\'{a}vala, Giulio Chiribella, and Howard Barnum for helpful discussions.

This research was supported in part by Perimeter Institute for Theoretical Physics. Research at Perimeter Institute is supported by the Government of Canada through the Department of Innovation, Science and Economic Development Canada and by the Province of Ontario through the Ministry of Research, Innovation and Science.\end{acknowledgments}

\bibliographystyle{plain}
\bibliography{CCF_bib}

\begin{thebibliography}{10}

\bibitem{ACGKM15}
Dorit Aharonov, Andr{\'e} Chailloux, Maor Ganz, Iordanis Kerenidis, and
  Lo{\"\i}ck Magnin.
\newblock A simpler proof of existence of quantum weak coin flipping with
  arbitrarily small bias.
\newblock {\em SIAM Journal of Computing}, 45(3):633--679.

\bibitem{ARW18}
Atul~Singh Arora, J\'{e}r\'{e}mie Roland, and Stephan Weis.
\newblock Quantum weak coin flipping.
\newblock {\em arXiv preprint arXiv:1811.02984}, 2018.

\bibitem{bae2016structure}
Joonwoo Bae, Dai-Gyoung Kim, and Leong-Chuan Kwek.
\newblock Structure of optimal state discrimination in generalized
  probabilistic theories.
\newblock {\em Entropy}, 18(2):39, 2016.

\bibitem{BCJRWY14}
Somshubhro Bandyopadhyay, Alessandro Cosentino, Nathaniel Johnston, Vincent
  Russo, John Watrous, and Nengkun Yu.
\newblock Limitations on separable measurements by convex optimization.
\newblock {\em IEEE Transactions on Information Theory}, 61(6):3593--3604,
  2015.

\bibitem{barnum2008nonclassicality}
Howard Barnum, Oscar~CO Dahlsten, Matthew Leifer, and Ben Toner.
\newblock Nonclassicality without entanglement enables bit commitment.
\newblock In {\em Information Theory Workshop, 2008. ITW'08. IEEE}, pages
  386--390. IEEE, 2008.

\bibitem{barnum2018oracles}
Howard Barnum, Ciar{\'a}n~M Lee, and John~H Selby.
\newblock Oracles and query lower bounds in generalised probabilistic theories.
\newblock {\em Foundations of physics}, 48(8):954--981, 2018.

\bibitem{barnum2011information}
Howard Barnum and Alexander Wilce.
\newblock Information processing in convex operational theories.
\newblock {\em Electronic Notes in Theoretical Computer Science}, 270(1):3--15,
  2011.

\bibitem{barrett2007information}
Jonathan Barrett.
\newblock Information processing in generalized probabilistic theories.
\newblock {\em Physical Review A}, 75(3):032304, 2007.

\bibitem{barrett2017computational}
Jonathan Barrett, Niel de~Beaudrap, Matty~J Hoban, and Ciar{\'a}n~M Lee.
\newblock The computational landscape of general physical theories.
\newblock {\em arXiv preprint arXiv:1702.08483}, 2017.

\bibitem{barrett2005no}
Jonathan Barrett, Lucien Hardy, and Adrian Kent.
\newblock No signaling and quantum key distribution.
\newblock {\em Physical review letters}, 95(1):010503, 2005.

\bibitem{Blu81}
Manuel Blum.
\newblock Coin flipping by telephone.
\newblock In Allen Gersho, editor, {\em Advances in Cryptology: A Report on
  CRYPTO 81, CRYPTO 81, IEEE Workshop on Communications Security, Santa
  Barbara, California, USA, August 24-26, 1981}, pages 11--15. U. C. Santa
  Barbara, Dept. of Elec. and Computer Eng., ECE Report No. 82-04, 1982, 1981.

\bibitem{BV}
Stephen Boyd and Lieven Vandenberghe.
\newblock {\em Convex Optimization}.
\newblock Cambridge University Press, 2004.

\bibitem{CK09}
Andr{\'e} Chailloux and Iordanis Kerenidis.
\newblock Optimal quantum strong coin flipping.
\newblock In {\em Proceedings of 50th IEEE Symposium on Foundations of Computer
  Science}, pages 527--533. IEEE Computer Society, 2009.

\bibitem{chiribella2010probabilistic}
Giulio Chiribella, Giacomo~Mauro D'Ariano, and Paolo Perinotti.
\newblock Probabilistic theories with purification.
\newblock {\em Physical Review A}, 81(6):062348, 2010.

\bibitem{davies1970operational}
E~Brian Davies and John~T Lewis.
\newblock An operational approach to quantum probability.
\newblock {\em Communications in Mathematical Physics}, 17(3):239--260, 1970.

\bibitem{fiorini2014generalized}
Samuel Fiorini, Serge Massar, Manas~K Patra, and Hans~Raj Tiwary.
\newblock Generalized probabilistic theories and conic extensions of polytopes.
\newblock {\em Journal of Physics A: Mathematical and Theoretical},
  48(2):025302, 2014.

\bibitem{garner2018interferometric}
Andrew~JP Garner.
\newblock Interferometric computation beyond quantum theory.
\newblock {\em Foundations of Physics}, 48(8):886--909, 2018.

\bibitem{GSU13}
Sevag Gharibian, Jamie Sikora, and Sarvagya Upadhyay.
\newblock {QMA} variants with polynomially many provers.
\newblock {\em Quantum Information \& Computation}, 13(1\&2):0135--0157, 2013.

\bibitem{GW07}
Gus Gutoski and John Watrous.
\newblock Toward a general theory of quantum games.
\newblock In {\em Proceedings of the Thirty-Ninth Annual ACM Symposium on
  Theory of Computing}, pages 565--574, New York, NY, USA, 2007. ACM.

\bibitem{hardy2001quantum}
Lucien Hardy.
\newblock Quantum theory from five reasonable axioms.
\newblock {\em arXiv preprint arXiv:0101012}, 2001.

\bibitem{hardy2011reformulating}
Lucien Hardy.
\newblock Reformulating and reconstructing quantum theory.
\newblock {\em arXiv preprint arXiv:1104.2066}, 2011.

\bibitem{JP17}
Anna Jen\v{c}ov\'{a} and Martin Pl\'{a}vala.
\newblock Conditions on the existence of maximally incompatible two-outcome
  measurements in general probabilistic theory.
\newblock {\em Physical Review A}, 96:022113, 2017.

\bibitem{Kit03}
Alexei Kitaev.
\newblock Quantum coin-flipping.
\newblock Unpublished result. Talk at the 6th Annual workshop on Quantum
  Information Processing (QIP 2003), 2002.

\bibitem{krumm2018quantum}
Marius Krumm and Markus~P Mueller.
\newblock Quantum computation is an island in theoryspace.
\newblock {\em arXiv preprint arXiv:1804.05736}, 2018.

\bibitem{lami2018ultimate}
Ludovico Lami, Carlos Palazuelos, and Andreas Winter.
\newblock Ultimate data hiding in quantum mechanics and beyond.
\newblock {\em Communications in Mathematical Physics}, 361(2):661--708, 2018.

\bibitem{LP15}
Monique Laurent and Teresa Piovesan.
\newblock Conic approach to quantum graph parameters using linear optimization
  over the completely positive semidefinite cone.
\newblock {\em Siam J. Optim.}, 25(4):2461--2493, 2015.

\bibitem{lee2015computation}
Ciar{\'a}n~M Lee and Jonathan Barrett.
\newblock Computation in generalised probabilisitic theories.
\newblock {\em New Journal of Physics}, 17(8):083001, 2015.

\bibitem{lee2016bounds}
Ciar{\'a}n~M Lee and Matty~J Hoban.
\newblock Bounds on the power of proofs and advice in general physical
  theories.
\newblock {\em Proc. R. Soc. A}, 472(2190):20160076, 2016.

\bibitem{lee2016deriving}
Ciar{\'a}n~M Lee and John~H Selby.
\newblock Deriving {G}rover's lower bound from simple physical principles.
\newblock {\em New Journal of Physics}, 18(9):093047, 2016.

\bibitem{lee2016generalised}
Ciar{\'a}n~M Lee and John~H Selby.
\newblock Generalised phase kick-back: the structure of computational
  algorithms from physical principles.
\newblock {\em New Journal of Physics}, 18(3):033023, 2016.

\bibitem{lee2017higher}
Ciar{\'a}n~M Lee and John~H Selby.
\newblock Higher-order interference in extensions of quantum theory.
\newblock {\em Foundations of Physics}, 47(1):89--112, 2017.

\bibitem{LC97a}
Hoi-Kwong Lo and Hoi~Fung Chau.
\newblock Why quantum bit commitment and ideal quantum coin tossing are
  impossible.
\newblock {\em Physica D: Nonlinear Phenomena}, 120(1--2):177--187, 1998.
\newblock 

\bibitem{Ludwig}
G.~Ludwig.
\newblock {\em An Axiomatic Basis of Quantum Mechanics. 1. Derivation of
  Hilbert Space}.
\newblock Springer-Verlag, 1985.

\bibitem{Mackey}
G.~W. Mackey.
\newblock {\em The mathematical foundations of quantum mechanics}.
\newblock W. A. Benjamin, New York, 1963.

\bibitem{Moc07}
Carlos Mochon.
\newblock Quantum weak coin flipping with arbitrarily small bias.
\newblock Available as arXiv.org e-Print quant-ph/0711.4114, 2007.

\bibitem{NST16}
Ashwin Nayak, Jamie Sikora, and Levent Tun\c{c}el.
\newblock A search for quantum coin-flipping protocols using optimization
  techniques.
\newblock {\em Mathematical Programming}, 156(1-2):581--613, 2016.

\bibitem{NST15}
Ashwin Nayak, Jamie Sikora, and Levent Tun{\c{c}}el.
\newblock Quantum and classical coin-flipping protocols based on bit-commitment
  and their point games.
\newblock Available as arXiv.org e-Print quant-ph/1504.04217, 2015.

\bibitem{Piron64}
C.~Piron.
\newblock Axiomatique quantique.
\newblock {\em Helvetia Physica Acta}, 37:439--468, 1964.

\bibitem{randall1970approach}
CH~Randall and DJ~Foulis.
\newblock An approach to empirical logic.
\newblock {\em The American Mathematical Monthly}, 77(4):363--374, 1970.

\bibitem{selby2018make}
John~H Selby and Jamie Sikora.
\newblock How to make unforgeable money in generalised probabilistic theories.
\newblock {\em Quantum}, 2:103, 2018.

\bibitem{Shapiro2009}
Alexander Shapiro.
\newblock Semi-infinite programming, duality, discretization and optimality
  conditions.
\newblock {\em Optimization}, 58(2):133--161, 2009.

\bibitem{sikora2018simple}
Jamie Sikora and John Selby.
\newblock Simple proof of the impossibility of bit commitment in generalized
  probabilistic theories using cone programming.
\newblock {\em Physical Review A}, 97(4):042302, 2018.

\bibitem{SW17}
Jamie Sikora and Antonios Varvitsiotis.
\newblock Linear conic formulations for two-party correlations and values of
  nonlocal games.
\newblock {\em Mathematical Programming}, 162(1-2):431--463, 2017.

\end{thebibliography}

\end{document}